\documentstyle[12pt]{article}
\begin{document}


\renewcommand{\thesection}{\arabic{section}}
\renewcommand{\theequation}{\arabic{equation}}
\renewcommand {\c}  {\'{c}}
\newcommand {\cc} {\v{c}}
\newcommand {\s}  {\v{s}}
\newcommand {\CC} {\v{C}}
\newcommand {\C}  {\'{C}}
\newcommand {\Z}  {\v{Z}}

\baselineskip=24pt

\begin{center}
{\bf   INTERACTING FAMILIES OF CALOGERO-TYPE PARTICLES AND SU(1,1) ALGEBRA}
 
\bigskip

S. MELJANAC$^{a}$ {\footnote{e-mail: meljanac@thphys.irb.hr}}, 
M. MILEKOVI\'{C} $^{b}$ {\footnote{e-mail: marijan@phy.hr }},
A. SAMSAROV$^{a}${ \footnote{e-mail:andjelo.samsarov@irb.hr}}
and M. STOJI\'{C}$^{a}${ \footnote{e-mail:marko.stojic@zg.hinet.hr  }}\\
$^{a}$ Rudjer Bo\v{s}kovi\'c Institute, Bijeni\v cka  c.54, HR-10002 Zagreb,
Croatia\\[3mm] 

$^{b}$ Theoretical Physics Department, Faculty of Science, P.O.B. 331,
 Bijeni\v{c}ka c.32,
\\ HR-10002 Zagreb, Croatia \\[3mm]

\bigskip

\end{center}
\setcounter{page}{1}
\bigskip

We study a one-dimensional  model with F interacting families of Calogero-type
particles. The model includes harmonic, two-body and three-body interactions. 
We emphasize the universal SU(1,1) structure of the model. 
We show how SU(1,1) generators for the whole system are composed of SU(1,1) 
generators of  arbitrary subsystems.
We find the exact eigenenergies corresponding to a class of the exact eigenstates of the 
F-family model.  By imposing the conditions for the absence of the three-body interaction, we
find  certain relations between the coupling constants. Finally, we establish  some relations 
of equivalence between  two  systems containing F families of Calogero-type particles.\\



PACS number(s): 03.65.Fd, 03.65.Sq, 05.30.Pr\\



\newpage

\section {Introduction}
Since its inception, the ordinary Calogero  model $^1$ continues to be of
interest for both physics and mathematics community$^2$.
The  model describes $N$ identical (single-species)  particles on 
the line which interact through an inverse-square two-body interaction and are subjected to 
a common confining harmonic force.  The inverse-square potential can be 
regarded as a pure statistical interaction $^3$ and the model maps to an ideal gas of particles 
obeying fractional Haldane statistics $^4$. 
The role of Haldane statistical parameters is played by a universal coupling constant in the
two-body interaction. However, in Haldane's formulation of statistics there is a 
possibility of having particles of different species with a mutual statistical coupling parameter depending
on the  species coupled. This suggests the generalization of the 
single-species Calogero model to the multispecies Calogero model. 
Distinguishabillity of the species can be introduced by allowing particles to have different masses 
and different couplings to each other. While the single-species Calogero model is completely 
solvable $^{1,5}$, very little is known about spectra and wave functions of the  
multispecies Calogero model $^6$.\\
Recently,  we used an operator method to analyse
a one-dimensional  multispecies Calogero model  with two- and three-body interactions $^7$.
We succeeded in finding  a class of, but not all, exact eigenstates and eigenenergies of the model  Hamiltonian.  
The analysis   relied heavily on the SU(1,1) algebraic structure of the Hamiltonian and once more stressed the 
importance of the conformal symmetry of the quantum singular oscillator $^8$. We were also able to
generalize the model of Ref.7 to  arbitrary dimensions $^9$.\\
In the present  Letter, which is in a sense a continuation of our
investigation of the ordinary  Calogero model $^{10}$ and the multispecies Calogero model $^{7,9}$, we 
turn to the important problem of interacting families of  Calogero-type particles in one dimension, a theme
which is already announced in $^7$. We  consider a model with a potential that generally includes harmonic, 
two-body and three-body interactions acting between particles belonging to different
families, as well as the interaction between particles belonging to the same family with the coupling constant that may
be different for different families. In Section 2 we prepare all necessary tools for handling the  problem of
interacting  families. We collect, without rederiving, the main results of the analysis of the 
one-dimensional multispecies Calogero model $^7$. In Section 3 we apply these results to the case of  two
interacting families of Calogero particles. We display the model Hamiltonian and find the ground state energy. We
construct generators of   SU(1,1) algebra  for interacting families and underline the
importance of the dilatation part of the algebra, i.e. generator $T_0$. Furthermore, by imposing the conditions for the
absence of the three-body interaction in the initial Hamiltonian, we find  certain relations between the coupling constants.
In Section 4 we extend  these results to   three and more interacting families of Calogero particles. We show that
the underlying SU(1,1) structure is universal, i.e. holds for an arbitrary number of families,  arbitrary masses of 
Calogero particles and arbitrary coupling constants. We particularly show how to obtain SU(1,1) generators of the whole
system from  SU(1,1) generators of arbitrary subsystem, i.e. we establish  composition rules
for  SU(1,1) generators. 
 We also find the 
${\em {exact}}$ eigenenergies corresponding to a class of the ${\em {exact}}$ eigenstates of the model with F
interacting families. We discuss the relations between the coupling constants in the case when a three-body
interaction vanishes. Finally, we establish some relations of equivalence between  two systems containing F
families. Section 5 is a short conclusion.



\section{A multispecies Calogero model: main results}
The model of Ref.7 is specified by masses of  particles,
$ m_{i}$, and the coupling constants $\omega$ and $\nu_{ij},\; i,j = 1,2,...,N$. The Hamiltonian
is 
$$
H(\omega) = -\frac{1}{2}\sum_{i=1}^{N}\frac{1}{m_{i}}\frac{{\partial}^{2}}
{\partial x_{i}^{2}} + \frac{{\omega}^{2}}{2}\sum_{i=1}^{N }m_{i} x_{i}^{2}
+ \frac{1}{4} \sum_{i \neq j }\frac{{\nu}_{ij}({\nu}_{ij}-1)}
{{(x_{i}-x_{j})}^{2}}(\frac{1}{m_{i}} + \frac{1}{m_{j}}) +
$$
\begin{equation}
+\frac{1}{2}\sum_{i \neq j, i \neq k, j \neq k
 }\frac{{\nu}_{ij}{\nu}_{jk}}
      {m_{j}(x_{j}-x_{i}) (x_{j}-x_{k})}.
\end{equation}
The ground state wave function is of the Calogero type:
\begin{equation}
\Psi_{0}(x_{1},...,x_{n})= \prod_{i<j}|x_{i}-x_{j}|^{\nu_{ij}}
e^{-\frac{\omega}{2}\sum_{i=1}^{N}
m_{i}x_{i}^{2}}\equiv
\Delta e^{-\frac{\omega}{2}\sum_{i=1}^{N}
m_{i}x_{i}^{2}}
\end{equation}
and the corresponding ground state energy is
\begin{equation}
E_0=\omega \epsilon_{0} = \omega ( \frac{N}{2} + \sum_{i < j}\nu_{ij} ).
\end{equation}
When all  couplings $\nu_{ij}$ are equal, Eq.(3) reduces to the well-known Calogero result 
$\epsilon_{0}=\frac{N}{2}+ \nu \frac{N (N-1)}{2}$.\\
After performing a similarity transformation
\[ \begin{array}{l}
\tilde{H}(\omega) = {\Delta}^{-1} H(\omega) \Delta ,\\
\tilde{\Psi} = {\Delta}^{-1} \Psi ,
\end{array}\]
one obtains a non-Hermitean Hamiltonian $\tilde{H}(\omega)$ with a hidden three-body interaction:
$$
\tilde{H}(\omega) = -\frac{1}{2}\sum_{i=1}^{N}\frac{1}{m_{i}}\frac{{\partial}^{2}}
{\partial x_{i}^{2}} + \frac{{\omega}^{2}}{2}\sum_{i=1}^{N }m_{i} x_{i}^{2}
- \frac{1}{2} \sum_{i \neq j }\frac{{\nu}_{ij}}
{(x_{i}-x_{j})}(\frac{1}{m_{i}} \frac{\partial}{{\partial} x_{i}}
 - \frac{1}{m_{j}} \frac{\partial}{{\partial} x_{j}})=
$$
\begin{equation}
={\omega}^{2}T_{+}-T_{-},
\end{equation}
where 
$$
T_{-} = -\tilde{H}(\omega = 0) ,\qquad T_{+} = \frac{1}{2}\sum_{i=1}^{N }m_{i} x_{i}^{2} ,
$$
\begin{equation}
T_{0} = \frac{1}{2} (\sum_{i=1}^{N} x_{i}\frac{\partial}{{\partial} x_{i}}
      + \epsilon_{0}). 
\end{equation}
The set of operators $\{T_{\pm},T_0\}$ satisfy the SU(1,1) algebra:
\begin{equation}
[T_{-},T_{+}] = 2T_{0}, \;\;\;\; [T_{0},T_{\pm}] = \pm T_{\pm}.
\end{equation}
Note that  $T_0$ serves as a dilatation operator. One can deduce that
$$
T_{0}\Delta = ( \frac {1}{2}\sum_{i<j} \nu_{ij} + \frac{\epsilon_{0}}{2} )\Delta
$$
\begin{equation}
T_{-}\Delta = 0.
\end{equation}
It is convenient to introduce the centre-of-mass coordinate
$ X = \frac{1}{M} \sum_{i=1}^{N}m_{i} x_{i} $ ( where 
 $ M = \sum_{i=1}^{N}m_{i}$ ) and relative coordinates  $ \xi_{i} = x_{i}-X $. In terms of these coordinates,  
the Hamiltonian $\tilde{H}(\omega)$, Eq.(4), separates
into parts which describe its centre-of-mass motion (CM) and its relative motion (R),
namely $\tilde{H}(\omega)=\tilde{H}(\omega)_{CM} + \tilde{H}(\omega)_{R}$. In the same way one can split the generators
 $ T_{\pm}$ and  $ T_{0}$ into the centre-of-mass and relative parts, i.e. $T_{\pm,0}=
T_{\pm,0_{(CM)}} + T_{\pm,0_{(R)}}$. \\
In the next section we  apply these results to the case of two interacting families.




\section{ Two interacting  families }
Let us consider two families, ${\cal {F}}_1$ and ${\cal {F}}_2$, of Calogero particles. 
The first one, denoted by 
$ {\cal {F}}_1= \{m_1, \nu_1, N_1\}$, is described by $ N_{1} $ particles of mass  $m_{1}$, the coupling constant
$\nu_{1}$ and the coordinates of the particles are $\{x_i\}=\{x_{1},x_{2},...,x_{N_{1}}\}$. 
Similarly, the second one, denoted by ${\cal {F}}_2 =\{m_2, \nu_2, N_2\}$, is described by $ N_{2} $ particles 
of mass  $m_{2}$,  the coupling constant $\nu_{2} $ and the coordinates of the particles are 
$\{ z_{\alpha}\}=\{z_{1},z_{2},...,z_{N_{2}}\}$. The interaction strength between the first and the second family 
is $\nu_{12} = \kappa$. \\
The full Hamiltonian now reads
\begin{equation}
H(\omega) = H_{1}(\omega) + H_{2}(\omega) + H_{int},
\end{equation}
where  $H_{int}$ is given by
$$
 H_{int} = \frac{1}{4} \sum_{i}^{N_{1}}\sum_{\alpha }^{N_{2}}
 \frac{\kappa(\kappa -1)}{(x_{i}-z_{\alpha})^{2}}\left(
\frac{1}{ m_1}  +  \frac{1} { m_2 } \right) +
$$

$$
+ \frac{1}{4}\sum_{i}^{N_{1}} \sum_{\alpha \neq \beta}^{N_{2}}
 \left( \frac{{\kappa}^{2}}
      {m_{1}(x_{i}-z_{\alpha}) (x_{i}-z_{\beta})}\right) +    
\frac{1}{2}\sum_{i}^{N_{1}} \sum_{\alpha \neq \beta}^{N_{2}} \left(\frac{\nu_2 \kappa}
      {m_{2} (z_{\alpha}-x_{i}) (z_{\alpha}-z_{\beta})} \right) + 
$$

\begin{equation}
 + \frac{1}{4}\sum_{i \neq j}^{N_{1}} \sum_{\alpha}^{N_{2}}
 \left( \frac{{\kappa}^{2}}
      {m_{2} (z_{\alpha} - x_{i}) (z_{\alpha} - x_{j})} \right )+  
\frac{1}{2}\sum_{i \neq j}^{N_{1}} \sum_{\alpha}^{N_{2}} \left( \frac{\nu_1 \kappa}
      {m_{1} (x_{i}-z_{\alpha}) (x_{i}-x_{j})} \right ), 
\end{equation}

and $H_1(\omega)$ ( $H_2(\omega)$ ) are Calogero Hamiltonians, Eq.(1), for the first and the second
family, respectively.\\
The corresponding ground state wave function of the Hamiltonian (8) is 
$$
\Psi_{0}(x_{1},...,x_{N_{1}},z_{1},...,z_{N_{2}}) = \prod_{i,\alpha}(x_{i}-z_{\alpha})^{\kappa}\Psi_{0,1}(x_{1},...,x_{N_{1}})
\Psi_{0,2}(z_{1},...,z_{N_{2}})
$$
\begin{equation}
\equiv\Delta_{12}\Psi_{0,1}(x_{1},...,x_{N_{1}})
\Psi_{0,2}(z_{1},...,z_{N_{2}}),
\end{equation}
where $\Psi_{0,1}$ and $\Psi_{0,2}$ are the Calogero ground states (when $\kappa =0$), Eq.(2),  for the
 families ${\cal {F}}_1$ and ${\cal {F}}_2$, respectively.\\
We can perform a similarity transformation with a  $\Delta_{1}\Delta_{2}$ part
of the full Jastrow prefactor $\Delta=\Delta_{1}\Delta_{2}\Delta_{12}$ in (8,10), to 
obtain
$$
{\Delta_{1}}^{-1}{\Delta_{2}}^{-1} H(\omega) \Delta_{1}\Delta_{2} = 
\tilde{H}_{1}(\omega) + \tilde{H}_{2}(\omega) + H_{int},
$$
\begin{equation}
{\Delta_{1}}^{-1}{\Delta_{2}}^{-1}\Psi_{0} =
 \prod_{i,\alpha}(x_{i}-z_{\alpha})^{\kappa}\tilde{\Psi}_{0,1}
\tilde{\Psi}_{0,2}.
\end{equation}
The ground state energy of the Hamiltonian (8) can be split into three terms:
\begin{equation}
\epsilon_{0} =  \epsilon_{0,1} + \epsilon_{0,2} + \kappa N_{1}N_{2},
\end{equation}
describing the ground state energies of  each family and 
the interaction between them, respectively. \\
For each family, one can define $SU(1,1)$ generators $T_{\pm}^{(I)}, T_{0}^{(I)}$, $I=1,2$. These
two sets of generators, i.e. the corresponding $SU(1,1)$ algebras, mutually commute. 
From the following  two relations:
$$
T_{0} = T_{0}^{(1)} + T_{0}^{(2)} +  \frac{1}{2} \kappa N_{1}N_{2},
$$
$$
T_{0}\Delta_{12} = \frac{1}{2}( \kappa N_{1}N_{2}  + \epsilon_{0})\Delta_{12},
$$
we find
\begin{equation}
(T_{0}^{(1)} + T_{0}^{(2)})  \Delta_{12} = \frac{\epsilon_{0}}{2}\Delta_{12}.
\end{equation} 
Furthermore, from Eq. (7) and after multiplication by ${\Delta_{1}}^{-1}{\Delta_{2}}^{-1}$ 
from the left, it follows that
\begin{equation}
T_{-}\Delta_{12} \equiv (T_{-}^{(1)} + T_{-}^{(2)} - H_{int})\Delta_{12} = 0.
\end{equation}
Note that  $ T_{+} = T_{+}^{(1)} + T_{+}^{(2)}$. 

As we have already shown in Ref.7, for the general $\nu_{ij}$ and $m_j$  the three-body interactions in the 
initial Hamiltonian (1) vanish  identically if the following conditions are satisfied for all triples of 
indices $i,j,k$: 
\begin{equation}
\frac{\nu_{ij} \nu_{jk}}{m_{j}} = \frac{\nu_{ji} \nu_{ik}}{m_{i}} =
\frac{\nu_{ik} \nu_{kj}}{m_{k}}.
\end{equation} 
In this case, the Hamiltonian contains the 
two-body interactions  (i.e. inverse-square interactions)   only. The unique solution of these conditions 
is $\nu_{ij}=\lambda \, m_i\, m_j$,  $\lambda $ being 
some universal constant.

In our two-family 
system this corresponds to the condition 
\begin{equation}
\nu_{ij} = \lambda m_{i} m_{j} = \kappa, \qquad    \forall \, i,j
\end{equation}
or explicitly
\begin{equation}
\nu_{1} = \lambda m_{1}^{2}, \qquad  \nu_{2} = \lambda m_{2}^{2},\qquad 
\nu_{12}= \kappa = \lambda  m_{1}m_{2}, 
\end{equation}
from which  it follows 
$$
 \nu_{1} \nu_{2} = {\kappa}^{2} ,  
$$
\begin{equation}
 \nu_{2} = {(\frac{m_{2}}{m_{1}})}^{2}\nu_{1}.
\end{equation}
Note that Eqs.(16-18) imply that the  couplings $\nu_{1}, \nu_{2}$ and $\kappa $ have to be simultaneously
positive, negative or zero.\\
The connection between the coupling constants $\{\nu_{1},\nu_{2},\kappa\}$, Eqs.(18), is ascribed to the  
weak-strong coupling duality in Ref.11, but it is  {\em  {de\, facto}} a simple consequence of the absence of the
three-body interaction in the starting Hamiltonian (1).
We also point out that all the above relations for $T_{0}^{(1)}+T_{0}^{(2)}$ 
and $T_{-}^{(1)} + T_{-}^{(2)}$ (Eqs.(13,14)) hold generally for  arbitrary masses
$m_{1},m_{2}$ and   arbitrary coupling constants  $ \nu_{1}, \nu_{2}, \kappa$, i.e. irrespectively of the 
presence/absence of the three-body interaction.\\




\section{Three and more interacting families, exact eigenstates and equivalences between models}
We extend the above analysis to the case of three families, 
${\cal {F}}_1$, ${\cal {F}}_2$ and ${\cal {F}}_3$, of Calogero particles.  
The families are characterized by mutually distinct numbers 
of particles, masses of particles and different coupling constants.
One can immediatelly generalize the results (13,14):
$$
\Delta = \Delta_{1}\Delta_{2}\Delta_{3}\Delta_{12}\Delta_{13}\Delta_{23},
$$
$$
T_{0} = T_{0}^{(1)} + T_{0}^{(2)} + T_{0}^{(3)} + \frac{1}{2}(\epsilon_{0}
- \epsilon_{0,1} - \epsilon_{0,2} - \epsilon_{0,3}),
$$
$$
T_{+} = T_{+}^{(1)} + T_{+}^{(2)} + T_{+}^{(3)},
$$
$$
T_{-} = T_{-}^{(1)} + T_{-}^{(2)} + T_{-}^{(3)} - H_{int},
$$
$$
(T_{0}^{(1)} + T_{0}^{(2)} + T_{0}^{(3)})\Delta_{12}\Delta_{13}\Delta_{23}
= \frac{\epsilon_{0}}{2}\Delta_{12}\Delta_{13}\Delta_{23},
$$
\begin{equation}
(T_{-}^{(1)} + T_{-}^{(2)} + T_{-}^{(3)} - H_{int})
 \Delta_{12}\Delta_{13}\Delta_{23} = 0.
\end{equation}
For the initial $ N $ -body   multispecies Calogero model we can write  composition laws for the
SU(1,1) generators:
$$
T_{0} = \sum_{i = 1}^{N} T_{0}^{(i)} + \frac{1}{2}\sum_{i<j}\nu_{ij}, 
$$
$$
T_{+} = \sum_{i = 1}^{N}T_{+}^{(i)}, \qquad T_{-} = \sum_{i = 1}^{N}T_{-}^{(i)}- H_{int},
$$
from which it follows
$$
T_{0}\Delta = \frac{1}{2}(\sum_{i<j}\nu_{ij} + \epsilon_{0})\Delta,
$$
$$
(\sum_{i = 1}^{N} T_{0}^{(i)})\Delta = \frac{\epsilon_{0}}{2} \Delta,
$$
\begin{equation}
(\sum_{i = 1}^{N}T_{-}^{(i)}  - H_{int})\Delta = 0.
\end{equation}
These relations are general, valid for an arbitrary number of families F
 (i.e. for an arbitrary partition of a multispecies Calogero model), and for 
an arbitrary choice of masses $ m_{i}$ and coupling constants $ \nu_{ij}$.

The infinite set of exact eigenstates of the Hamiltonian (1) can be 
constructed by applying ladder operators 
\begin{equation}
  A_{1}^{\pm} = \frac{1}{\sqrt{2}}(\sqrt{M\omega} X \mp
\frac{1}{\sqrt{M\omega}} \frac{\partial}{\partial X})
\end{equation}
 and 
\begin{equation}
 B_{2}^{\pm}= \frac{1}{2}(\omega T_{+} + \frac{T_{-}}{\omega}) \mp T_{0}- \frac{1}{2}{A_{1}^{\pm}}^{2}
\end{equation}
 to the vacuum 
$$
\tilde{\Psi}_{0}(x_1,x_2,\cdots x_N)
=\tilde{\Psi}_{0}(X)\tilde{\Psi}_{0}(\xi_1, \xi_2 \cdots \xi_N)=
e^{-\frac{M\omega}{2}X^2} e^{-\frac{\omega}{2}\sum^{N}_{i=1} m_i \xi_i^2} .
$$
The exact eigenstates (corresponding to the center-of-mass states and global dilatation states)  are
\begin{equation}
\tilde{\Psi}_{n_{1}n_{2}} = {A_{1}^{+}}^{n_{1}}{B_{2}^{+}}^{n_{2}},
\tilde{\Psi}_{0}. \qquad n_{1}, n_{2} = 0,1,2,,,
\end{equation}
The exact eigenenergies corresponding to these states are ($I,J=1,2,...F$)
$$
E_{n_{1}n_{2}}= \omega(n_{1} + 2n_{2} + \epsilon_{0}),
$$
$$
\epsilon_0=\sum_{I=1}^{F} \epsilon_{0,I} + \sum_{I<J; I,J=1 }^{F} \epsilon_{0,IJ},
$$
$$
\epsilon_{0,I}=  \frac{N_I}{2}   + \nu_I \frac{N_I (N_I-1)}{2} ,
$$
\begin{equation}
\epsilon_{0,IJ}=\nu_{IJ}\, N_I \, N_J .
\end{equation}
In the special case, when there is no  three-body interaction (i.e. relations (15) are satisfied), 
we can identify
$$
\nu_I = \lambda m_I^2, \qquad \forall I=1,2...F,
$$
\begin{equation}
\nu_{IJ}=\lambda m_I m_J , \qquad \forall I,J=1,2...F.
\end{equation}
Since the masses are positive, the couplings $\nu_I$ and $\nu_{IJ}$ have the same sign, depending on the sign of
the free parameter $\lambda$.

Now we establish some relations of equivalence between the  two systems containing F families of 
Calogero particles.

{\bf {Case 1.}} {\em {Complete equivalence of the two systems.}}\\
Let  $ {\em {S}}=\{\omega, m_I, \nu_I, \nu_{IJ}, N_I\}$ and ${\em {S'}}=\{\omega, m'_I, \nu'_I, \nu'_{IJ}, N'_I\}$
be two Calogero systems with  F families. We call them completely equivalent if
$$
\epsilon_{0,I}=\epsilon'_{0,I}, \qquad \epsilon_{0,IJ}=\epsilon'_{0,IJ}, \qquad N_I=N'_I.
$$
These  conditions imply 
\begin{equation}
\nu_I=\nu'_I, \qquad \nu_{IJ}=\nu'_{IJ}.
\end{equation}

{\bf {Case 2.}} {\em {Partial equivalence of the two systems.}}\\
We call the two systems $ {\em {S}}$ and ${\em {S'}}$ partially equivalent if
$$
\epsilon_{0}=\epsilon'_{0} ,
$$ 
while the number of particles, $N$ and $N'$, may be the same or different. For example, in the case of one-family
systems ($F=1$) and $N\neq N'$, the above  condition implies that
\begin{equation}
N + \nu N (N-1) =N' + \nu' N' (N'-1).
\end{equation}

{\bf {Case 3.}} {\em {Special case: the single system.}}\\
Consider a single system with F families of  Calogero particles. 
We can demand that  
$$ 
\epsilon_{0,1}=\epsilon_{0,I}, \qquad \epsilon_{0,12}=\epsilon_{0,IJ}, \qquad \forall I,J=1,2...F.
$$
In the case of the two-family system ($F=2$), we have
\begin{equation}
N_1 + \nu_1 N_1 (N_1-1) =N_2 + \nu_2 N_2 (N_2-1).
\end{equation}
We obtain very  interesting relations between the couplings $\nu_1$ and $ \nu_2$
if we impose the strong-weak duality  condition on the couplings, namely $\nu_1 \nu_2=1$.
(We fix $\kappa^2 =1$ in Eq.(18)). \\ 
The quadratic equation (28) then has two solutions:
$$
(i)  \qquad \nu_1=\frac{ N_2 - 1 }{N_1 - 1 } > 0, 
$$
$$
(ii) \qquad \nu_1= - ( \frac{N_2}{N_1} )  < 0.
$$
Their physical implications are summarized in  Table 1.

Table 1.
$$
\begin{tabular}{ccccc} \hline
$\nu_1$  & $\kappa$ & $\lambda$ & $\epsilon_0$ & Comments \\ \\ \hline

$\frac{ N_2 -1 }{ N_1-1 }$ &  +1  &  $\lambda >0 $   & $2 N_1 N_2 > 0$ & Physical solution, no three-body
                                                                            interaction. \\ \\ 

$- ( \frac{N_2}{N_1} ) $ &  +1 & $ -$ & $N_1 + N_2 > 0$ & Physical solution, with a three-body interaction.\\ \\ 

$\frac{ N_2 -1 }{ N_1-1 }$ &  - 1  & $ - $   & $ 0 $ & Unphysical solution,  with a three-body interaction. \\ \\

$- ( \frac{N_2}{N_1} ) $ &  -1 & $\lambda < 0 $& $ N_1 + N_2 - 2 N_1 N_2 < 0$ & Unphysical solution, no three-body 
                                                                                             interaction. \\ \\ \hline

\end{tabular}
$$
A few remarks are in order. \\
{\em {Remark 1}}. Notice that  solution $(i)$ requires $N_{1,2} \geq 2$. \\
{\em {Remark 2}}. If the generalized strong-weak duality condition is imposed, i.e. 
$\nu_I \nu_J =1$, ($I \neq J=1,2, ...F $), then it follows that it can be satisfied  for $F=2$ only.\\
{\em {Remark 3}}. In Refs.7 and 10, we showed that there existed a critical point $\epsilon_{0R}=0$ at which the
system described by $\tilde{H}(\omega)_R$ collapsed completely, i.e. the relative momenta, the relative energy and the
relative coordinates were all zero at this critical point.  The ground state was a
square-integrable function only for $\epsilon_{0R}>0$. This is the reason why we ascribe the term 'unphysical'
 to the  last solutions in Table 1.



\section {Conclusion}
In this Letter we have studied the most general Calogero model on the line with a three-body interaction possessing 
an arbitrary number of mutually interacting families of  Calogero particles.  We have found the 
exact eigenenergies corresponding to a class of the exact eigenstates of the model. We have established  
relations of equivalence between two systems with F families which imply a certain connection between the coupling
constants. Particularly interesting appear the  relations between the coupling constants in the single system with F
families of  Calogero particles when a strong-weak duality condition is imposed.
We have  paid special attention to the  SU(1,1) structure
of the model. We have found  certain relations between SU(1,1) generators that are universal for all choices of masses
and coupling constants.  We particularly show how to obtain SU(1,1) generators of the whole
system from  SU(1,1) generators of arbitrary subsystem. Moreover, the same relations are valid for an arbitrary number
 of dimensions and for all
potentials that behave as a kinetic energy term under the dilatation represented by the generator $T_0$. 
There is  only one difference between one and higher dimensions.  In the case of one dimension, one can exclude
the three-body  interaction between particles from the beginning, while there is no known way how to do this in dimensions
higher than one. Our results  can also be  extended to other systems with the underlying conformal or superconformal 
symmetry $^{12}$.

\bigskip
\bigskip



{\bf Acknowledgment}\\
This work was supported by the Ministry of Science and Technology of the Republic of Croatia under 
contracts No. 0098003 and No. 0119261.

\newpage



\end{document}